\documentclass[aps,twocolumn,showpacs,superscriptaddress,citeautoscript,reprint,prb,floatfix]{revtex4-1}

\usepackage{bm}
\usepackage{graphicx}
\usepackage{amsmath}
\usepackage{amsfonts}
\usepackage{bbold}
\usepackage{amssymb}
\usepackage{siunitx}
\usepackage{color}

\newcommand{\revision}[1]{\textcolor{black}{{#1}}}

\begin{document}

\title{Floquet engineering of Dirac cones on the surface of a topological insulator}

\author{A. D\'{i}az-Fern\'{a}ndez}

\affiliation{GISC, Departamento de F\'{\i}sica de Materiales, Universidad Complutense, E--28040 Madrid, Spain}

\author{E. D\'{\i}az}

\affiliation{GISC, Departamento de F\'{\i}sica de Materiales, Universidad Complutense, E--28040 Madrid, Spain}

\author{A. G\'{o}mez-Le\'{o}n}

\affiliation{Instituto de Ciencia de Materiales de Madrid (ICMM-CSIC), Cantoblanco, E--28049 Madrid, Spain}

\author{G. Platero}

\affiliation{Instituto de Ciencia de Materiales de Madrid (ICMM-CSIC), Cantoblanco, E--28049 Madrid, Spain}

\author{F. Dom\'{i}nguez-Adame}

\affiliation{GISC, Departamento de F\'{\i}sica de Materiales, Universidad Complutense, E--28040 Madrid, Spain}

\begin{abstract}

We propose to Floquet-engineer Dirac cones at the surface of a three-dimensional topological insulator. We show that a large tunability of the Fermi velocity can be achieved as a function of the polarization, direction and amplitude of the driving field. Using this external control, the Dirac cones in the quasienergy spectrum may become elliptic or massive, in accordance to experimental evidences. These results help us to understand the interplay of surface states and external ac driving fields in topological insulators. 
In our work we use the full Hamiltonian for the three-dimensional system instead of effective surface Hamiltonians, which are usually considered in the literature. Our findings  show that the Dirac cones in the quasienergy spectrum remain robust even in the presence of bulk states and, therefore, they validate the usage of effective surface Hamiltonians to explore the properties of Floquet-driven topological boundaries. Furthermore, our model allows us to introduce new out-of-plane field configurations, which cannot be accounted for by effective surface Hamiltonians. 

\end{abstract}


\maketitle

\section{Introduction}

During the last two decades, new Dirac materials such as topological insulators, graphene and other carbon-based materials have emerged. These are foreseen to surpass the reach of semiconductors. Apart from their robustness to defects, stemming either from topological protection or symmetry, their linear dispersion is very much like that of photons, except for their quantum statistics and their much lower velocities. Different mechanisms have been put forward to modify the properties of these cones. For instance, breaking time-reversal symmetry in graphene leads to the quantum anomalous Hall effect, a system introduced by Haldane~\cite{Haldane88} in the 1980's and experimentally realized very recently using ultracold atoms~\cite{Jotzu14}. In this case, bandgaps open up in the otherwise gapless spectrum and the system becomes a topological insulator that can host chiral edge states. Other alternatives put their emphasis towards modifying the Fermi velocity~\cite{Li09,Trambly10,Hicks11,Hwang12,Elias11,Miao13}, a crucial parameter in quantum transport~\cite{Lima16}. As an example, applying static, uniform electric and magnetic fields to three-dimensional topological insulators such as Bi$_2$Se$_3$ widen the cone elliptically, so that the Fermi velocity is reduced in an anisotropic fashion~\cite{Diaz-Fernandez17a,Diaz-Fernandez17b,Diaz-Fernandez17c,Diaz-Fernandez18}. 
 
Remarkably, however, the use of periodic drivings is dramatically expanding the possibilities in these Dirac materials. Indeed, examples are now not only found in solid state systems~\cite{Lindner11}, but also in photonics~\cite{Rechtsman13} or even acoustics~\cite{Fleury16}. All these make use of what is known as Floquet's theorem. Although the words are now mainstream in the scientific community, Floquet's theorem is most well-known in its real space version, that is, Bloch's theorem. Indeed, the discrete periodicity of a lattice in real space leads to the concepts of energy bands and Brillouin zones. The same knowledge can be directly transferred to the domain of discretely time-periodic systems. In this case, there are quasienergies, in analogy to the quasimomentum of Bloch's theory, Floquet-Brillouin zones, and so forth~\cite{Grifoni98,Platero04,GomezLeon13a, Cayssol13}. In regard to the study of Dirac cones on the surface of a topological insulator, it has been experimentally observed~\cite{Wang13} and theoretically discussed~\cite{Kitagawa11} that these can be notably altered by applying time-periodic \emph{in-plane\/} fields. 
 
In our work we use a model that was introduced in a series of seminal papers started off by Volkov and Pankratov~\cite{Volkov85,Korenman87,Agassi88,Pankratov90,Adame94} in the 1980's and that is regaining very much interest lately in the context of surface states in three-dimensional topological insulators~\cite{Tchoumakov17,Inhofer17}. We will show that different orientations of the applied field with respect to the surface, as well as different polarizations, lead to a variety of situations. It is worth noticing that it has been already shown for example that an in-plane, circularly polarized field leads to gap openings~\cite{Wang13}, \revision{a feature that has been also observed in graphene}~\cite{Kibis10,Kitagawa11,Kibis16,Kibis17}. 
\revision{In view of previous studies based on graphene}~\cite{Syzranov13,Agarwala16}, \revision{phosphorene}~\cite{Iurov17}, \revision{$\alpha-$T$_3$ materials}~\cite{Iurov19} \revision{and three-dimensional topological insulators}~\cite{Yudin16}, other in-plane configurations are expected to preserve the Dirac point, isotropically or anisotropically widening the Dirac cone. In this paper, we will confirm these results on the surface of a topological insulator and, furthermore, we will extend previous studies with a detailed characterization of i) Dirac cones on the topological surface when a time-periodic out-of-plane field is applied and ii) the dependence of the main magnitudes of interest, the Fermi velocity and the gap, on the field parameters. The aforementioned references focus on the effective Hamiltonian for the surface states, performing perturbation theory in the high-frequency limit. \revision{In Ref.}~\onlinecite{Iurov13}, \revision{the three-dimensional Hamiltonian is mentioned to comment on the gap openings that occur when considering thin films of topological insulators, although the interplay between bulk and surface states is not discussed.} In our case, we will consider the high frequency limit as well, although we shall consider \revision{throughout the whole paper} the full Hamiltonian of the topological boundary. The usage of the full Hamiltonian allows us to observe the interplay with bulk states, which are not accessible to the effective surface Hamiltonian.

\section{Topological Boundary}

Topological materials can be characterized by an integer that is related to discrete symmetries of the bulk. For instance, Chern insulators are characterized by nonzero Chern numbers that arise when breaking time-reversal symmetry~\cite{Hasan10}. The words \emph{topological insulators} are usually reserved to systems that do preserve time-reversal symmetry and are generally classified according to $\mathbb{Z}_2$ indices~\cite{Hasan10}. Since an integer cannot change continuously, if two insulators of different topological index are placed together, at their interface there must be gapless modes. Otherwise, both systems would be connectable in a continuous way, implying that their invariants must be the same. As a result, the edge in two-dimensions or surface in three-dimensions formed in the contact region between these materials is known as a topological boundary~\cite{Zhang12}. In this section, we will consider Bi$_2$Se$_3$, an outstanding candidate for the foreseen applications of these materials. This is in part because of its wide bandgap, which allows it to perform even at room temperature~\cite{Hasan10}, but also because it is a well-known thermoelectric material and its experimental growth and characterization are now almost routine. The model is based on $\bm{k}\cdot\bm{p}$ theory and it was put forward by Volkov and Pankratov in the 1980's and it is currently recapturing very much attention~\cite{Volkov85,Tchoumakov17,Inhofer17}.

In the orbital-spin basis, $\left\{\tau,\sigma\right\}$, the bulk Hamiltonian of Bi$_2$Se$_3$ is a Dirac-like Hamiltonian of the form~\cite{Zhang12,Tchoumakov17}
\begin{equation}
    H = \bm{\alpha}\cdot\left(\bm{k}+\bm{A}\right) + \beta \ ,
    \label{eq:01}
\end{equation}
where $\bm{\alpha}=(\alpha_x,\alpha_y,\alpha_z)$, being $\alpha_j=\tau_x\otimes\sigma_j$ with $j=x,y,z$, $\beta=\tau_z\otimes\mathbb{1}_2$ the Dirac matrices, $\tau_j$ and $\sigma_j$ the Pauli matrices and $\mathbb{1}_d$ the $d$-dimensional identity matrix. Hereafter we will set $\hbar=1$. Energies will be expressed in units of half the bulk gap, $\Delta=E_{G}/2$, and there is a natural length scale, $d=v_{F}/\Delta$, where $v_{F}$ is the Fermi velocity. Momentum ${\bm k}$ is therefore expressed in units of $1/d$ and the vector potential ${\bm A}$ in units of $1/ed$, where $e$ is the elementary charge. In Bi$_2$Se$_3$, $E_{G}\simeq \SI{350}{\milli\electronvolt}$ and $v_{F}\simeq\SI{25}{\electronvolt\nano\meter}$, leading to $d\simeq \SI{2}{\nano\meter}$. The spectrum of this Hamiltonian in the absence of driving fields, that is, if $\bm{A}=0$, corresponds to that of a massive Dirac fermion, $E(k)=\pm\sqrt{1+k^2}$ with $k=|{\bm k}|$, the two bands being doubly degenerate. In addition, the eigenstates of equation~(\ref{eq:01}) are characterized by a non-vanishing $\mathbb{Z}_2$ topological invariant given by $\nu=\mathrm{sgn}(\Delta)$~\cite{Zhang12}.

In this case, a topological boundary is formed by introducing a position-dependent gap. This allows the system to have opposite bandgaps on each side of the boundary, changing the value of the $\mathbb{Z}_2$ topological invariant. The actual meaning of this is that the gap, defined as a difference between band edges of a certain orbital character or parity, changes sign because of a band inversion. Therefore, if we form a boundary between two systems with opposite bandgaps described by this Hamiltonian, there will be a change in the topological index and, as a result, there will be gapless modes at the boundary. Indeed, in the simplest case of a symmetric junction, the Hamiltonian above is modified to
\begin{equation}
    H = \bm{\alpha}\cdot\left(\bm{k}+\bm{A}\right) + \beta\mathrm{sgn}(z) \ ,
    \label{eq:02}
\end{equation}
where $z$ is the coordinate along the growth direction. It is not particularly difficult to show that in this case there is a midgap state, localized at the boundary with a localization length of $d$ and extended along the boundary plane. The dispersion in that plane is that of a single Dirac cone, $E(k_\bot)=\pm k_\bot$. Here the subscript $\bot$ indicates that the $z$ component of a vector is zero. These cones can coexist with doubly degenerate massive Volkov-Pankratov states if the interface is sufficiently smooth~\cite{Tchoumakov17}, in contrast to the sharp interface considered in this paper. Interestingly enough, applying static external electric and magnetic fields, it is possible to anisotropically widen the cone, therefore leading to an effective reduction of the Fermi velocity~\cite{Diaz-Fernandez17a,Diaz-Fernandez17b,Diaz-Fernandez17c,Diaz-Fernandez18}. In fact, it is straightforward to obtain analytic expressions for small enough fields. For instance, it can be explicitly shown that the Fermi velocity decreases with the applied field in a quadratic manner~\cite{Diaz-Fernandez17a}. As we will show below, specific configurations of the irradiated samples share this exact same characteristic. 

\section{Floquet engineering}

If we apply a time-periodic driving to the system instead of static fields, a wider range of situations occur. It is known from the use of surface effective Hamiltonians that a circularly polarized field will lead to gap openings~\cite{Kitagawa11}. However, only in the case of graphene it has been shown that the Dirac cones become strongly anisotropic in the case of linearly polarized fields~\cite{Syzranov13}. In the following, we shall show that these two features arise when the topological boundary Hamiltonian above is considered. More importantly, it allows us to consider out-of-plane configurations, which are not accessible to the aforementioned surface effective Hamiltonians. 

Hereafter we consider the system size to be small enough so as to ignore any spatial dependence of the field~\cite{Usaj14}. In that case, we can choose the vector potential components to be
\begin{equation}
    A_j(t) = a_je^{i\omega t} + a_{j}^{*}e^{-i\omega t} \ ,
    \label{eq:03}
\end{equation}
where $a_j=(f_j/2\omega)\exp(i\theta_j)$. Here, $f_j$ are the components of the electric field, $\bm{F}(t)=-\partial_t\bm{A}(t)$, measured in units of $\Delta/ed$, $\omega$ is the driving frequency measured in units of $\Delta$, and $\theta_j$ are phases which can be tuned to obtain different polarizations. The symmetries of this problem allow us to introduce three good quantum numbers. On the one hand, as a consequence of continuous translational symmetry in the $XY$ plane, the in-plane momenta $\bm{k}=(k_x,k_y,0)$ are good quantum numbers. On the other hand, discrete translational symmetry in time leads to the quasienergies, a central concept in Floquet theory. The discreteness of this symmetry restricts the quasienergies to the first Floquet-Brillouin zone, $\varepsilon\in[-\omega/2,\omega/2]$, very much like the quasimomentum in a lattice. All in all, it is possible to express the envelope function upon which the Hamiltonian acts as follows
\begin{equation}
    \bm{\Psi}(\bm{r},t) = e^{-i\varepsilon t}e^{i\bm{k}\cdot\bm{r}}\bm{\Phi}(z,t) \ ,
    \label{eq:04}
\end{equation}
where $\bm{\Phi}(z,t)=\bm{\Phi}(z,t+T)$ and $T=2\pi/\omega$. Notice that the problem is now very much simplified. Indeed, there is now only a $z$-dependence and the problem is reduced to a unit cell of size $T$ along the time axis. Hence, the equation to be solved for $\bm{\Phi}(z,t)$ is given by
\begin{equation}
    \varepsilon\bm{\Phi}(\bm{r},t) = \big(H-i\partial_t\big)\bm{\Phi}(\bm{r},t) \ .
    \label{eq:05}
\end{equation}
Taking advantage of the periodicity of $\bm{\Phi}(z,t)$, we can Fourier expand
\begin{equation}
    \bm{\Phi}(z,t) = \sum_{l=-\infty}^{\infty} \bm{\varphi}_{l}(z)e^{-il\omega t} \ .
    \label{eq:06}
\end{equation}
Indeed, it is possible to find straightforwardly an equation for the Fourier components
\begin{equation}
    \varepsilon\bm{\varphi}_{l}(z) = \big[\bm{\alpha}\cdot\bm{k}+\beta\mathrm{sgn}(z)-l\omega\mathbb{1}_4\big]\bm{\varphi}_{l}(z) 
    + J\bm{\varphi}_{l+1}(z)+J^{\dagger}\bm{\varphi}_{l-1}(z) \ ,
\label{eq:08}
\end{equation}
where $J=\bm{\alpha}\cdot\bm{a}$, with $\bm{a}$ a vector whose components are the previously defined $a_j$'s. 

Several comments are in order before continuing. The first is that, if we remove the field by setting $J=0$, the result is similar to that of free electrons when we imagine folding the energies by artificially introducing Brillouin zones. That is, the spectrum in the first Floquet-Brillouin zone can be obtained by repeatedly folding the spectrum for the driving-free case. For instance, for the topological boundary, the first Floquet-Brillouin zone displays evenly spaced cones, where the separation between consecutive Dirac points is $\omega$. Similar to the free electrons' case where the presence of a potential may open up energy gaps at the edges of the Brillouin zone, the presence of a non-zero $J$ leads to avoided crossings at the edges of the Floquet-Brillouin zone~\cite{Wang13}. The second point to notice is that, in the absence of boundary, that is, if there is no $z$-dependence, the equation is similar to that of a nearest-neighbors tight-binding problem with four orbitals per lattice site and a site dependent on-site energy due to the factor $l\omega$. This case is readily solved by diagonalization of a block-tridiagonal matrix. Third, time-reversal symmetry is broken only if a circularly polarized laser field is applied. Indeed, if the laser is linearly polarized, we can always choose the phases to be zero and, as a result, $J$ would be Hermitian. Alternatively, we can write $J=\exp(i\theta)\tilde{J}$, where $\tilde{J}$ is Hermitian and the phase factor can be eliminated via a gauge transformation of the form $\bm{\varphi}_l\to \exp[-i(l-1)\theta]\bm{\varphi}_l$. Therefore, it is expected that a circularly polarized field will lead to gap openings, whereas a linearly polarized field will not. We shall see in the following that this is indeed the case when the field is properly oriented.

In order to make further progress in the topological boundary case, it becomes necessary to discretize the Hamiltonian in the $z$-direction. Following Ref.~\citenum{Diaz14}, it is convenient to perform an alternate sampling of the components of $\bm{\varphi}_l$. That is, we will consider the discrete lattice in the $z$-direction to be composed of two sublattices, one for the even sites and one for the odd sites. The first and fourth components of $\bm{\varphi}_l$ will be sampled in the even sites, whereas the second and third components in the odd ones. This is explained in further detail in the Supplementary Information. Generally speaking in our numerical approach the system is placed in a box of size $L>1$ in the $z$-direction such that the real space variable is discretized in a one-dimensional lattice. In addition, we imposed a cutoff to the sideband or Fourier index and took great care to separate the bulk to the surface physics. Indeed, since we have placed the system in a box, the bands in the continuum will form subbands and these will enter the first Floquet-Brillouin zone upon band folding. In order to establish whether the Dirac state remains localized at the boundary despite the application of the external field, a careful analysis of the effect of the box size and the discretization step was performed. Indeed, if the box size (discretization step) is increased (decreased), more bulk quasienergies within the first Floquet-Brillouin zone will arise. However, if upon doing so the Dirac state remains unaltered, then we will conclude that it is localized at the boundary and therefore, it is well separated from the bulk states. Thus in such a case no hybridization between the Dirac and the bulk states is demonstrated. As a final remark, let us stress that our final objective is to characterize the reshaping of the Dirac cones under a small field perturbation. Thus, two requirements are fulfilled in our study: i) driving frequencies are larger than any other energy scale of the problem and ii)  driving amplitudes are small ($f/\omega<1$) so that the perturbations $a_j$ are also small.

\section{Results and Discussion}

In this section, we shall discuss four different cases of orientation and polarization of the incident field: in- and out-of-plane, linearly and circularly polarized fields. 
Hereafter, non-zero field amplitudes will be the same in all directions and we shall denote them collectively by $f$.

Before considering every case in detail, we would like to comment some common features. First, in all cases the resulting Dirac cones, or the double-sheeted hyperboloid in the case of a circularly polarized in-plane field, widen isotropically or anisotropically, depending on the orientation and polarization, upon increasing the field. Second, we will see below that it is possible to perfectly fit the change in the Fermi velocity as a quadratic function of the form $v_{F}(f)/v_{F}(0)=1-\gamma (f/\omega)^2$, where $\gamma$ depends on the orientation, the polarization and the frequency of the driving field.

\subsection{In-plane fields}

Having said that, let us start analyzing those orientations that have already been reported in the literature for graphene and for effective surface Hamiltonians of topological insulators~\cite{Kitagawa11,Syzranov13}. That is, we consider in-plane fields with linear and circular polarizations. 

\begin{figure}[htb]
\centerline{\includegraphics[width=0.95\columnwidth]{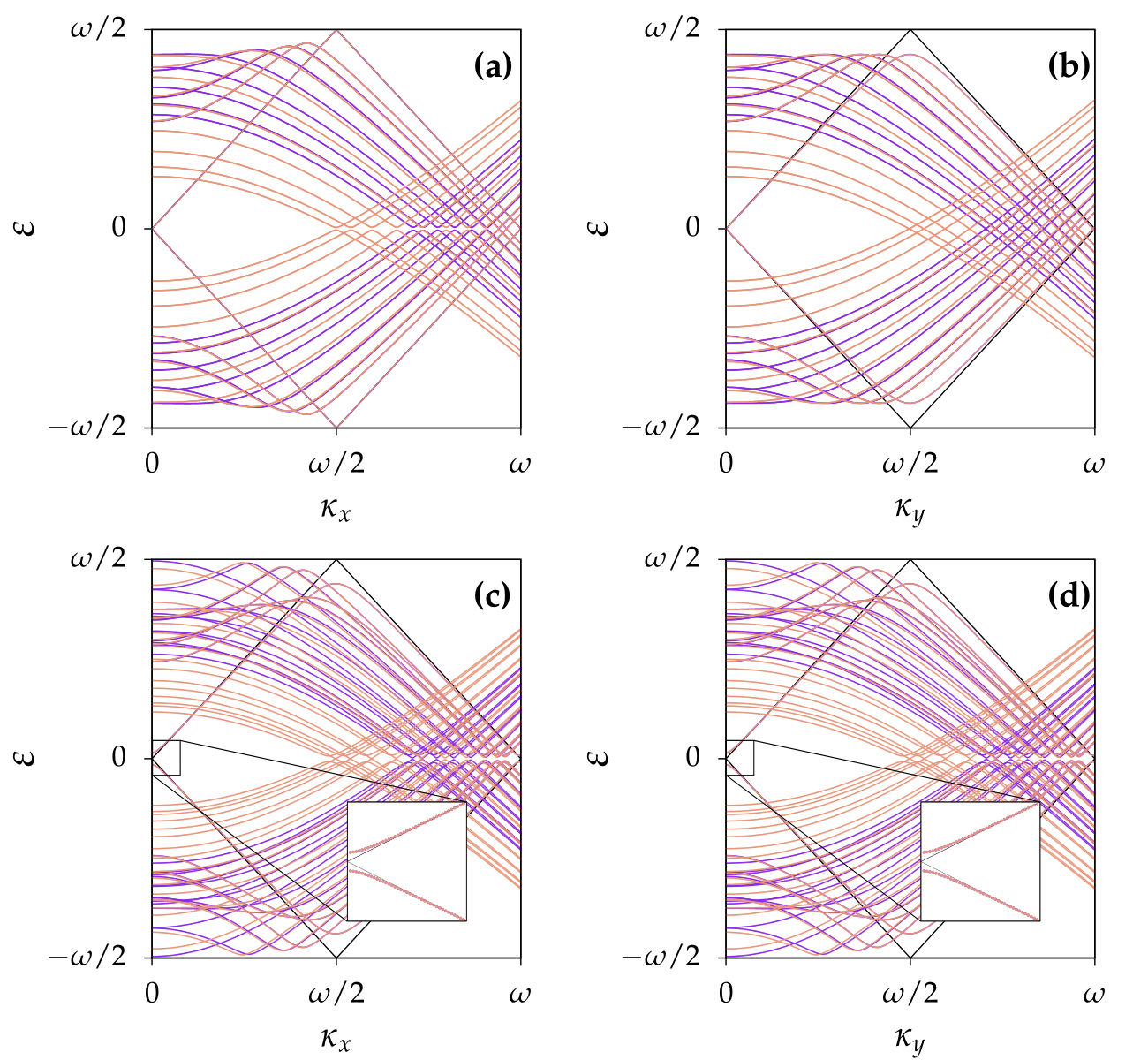}}
\caption{Quasienergy spectra for in-plane fields as a function of the momentum $\kappa_x$ and $\kappa_y$. In all cases, $\omega=4$ and $f=2$. Black lines indicate the Dirac cone replicas in the absence of perturbation (there would be bulk states as well). Blue and orange lines correspond to lattice spacings of $0.375$ and $0.300$, respectively. In all figures, avoided crossings occur at the Brillouin zone edges for the bulk states. Panels~(a) and~(b) correspond to linear polarization with the field along the $X$ direction. 
Panels (c) and (d) correspond to circular polarization. A gap opens up at the Dirac point and a widened massive dispersion occurs, as observed in the inset.}
\label{fig:inspectra}
\end{figure}

First, in order to accurately assess the localization of the surface state, we will perform the numerical calculations for a box of size $L = 3$ and two different grid spacing of $0.375$ and $0.300$. Additionally, we set $\omega = 4$ and a cutoff to the sideband index at $l = 3$. Figure~\ref{fig:inspectra} shows the resulting quasienergy spectra in the linearly and the circularly polarized cases.

There are a number of features to observe in this figure. First, upon increasing the discretization step, the number of subbands in the bulk states increases. Blue corresponds to the smaller discretizations step. However, the Dirac state is unchanged upon increasing the step and the dispersions overlap. Next, we can observe that there are avoided crossings at the edges of the Floquet-Brillouin zone, except for the Dirac state in Fig.~\ref{fig:inspectra}(a). According to Ref.~\onlinecite{Farrell16} this can be understood from the fact that the perturbation $f_x\alpha_x$ commutes with  $\boldsymbol{\alpha}_\bot\cdot \boldsymbol{\kappa}$ when $\kappa_y = 0$, whereas it does not when $\kappa_x = 0$. Hence, the perturbation does not couple the Dirac sidebands in the first case. Another observation that can be made is the fact that, due to the need to perform avoided crossings at the edges of the Floquet-Brillouin zone, the slope of the Dirac spectrum is reduced for low momenta. Hence, the dispersion is an anisotropic cone, widening in the direction perpendicular to the perturbation. This result is similar to what has been found for graphene in Ref.~\onlinecite{Syzranov13}. In our case, however, we are proving that this also occurs in topological insulators, despite the presence of bulk states. Hence, our results confirm that an effective surface Hamiltonian can be used to model the physics discussed here, since the bulk states and the surface states remain uncoupled. In the following we will analyze in  detail the reshaping of Dirac cones in topological boundaries under in-plane fields.

\begin{figure}[htb]
\centerline{\includegraphics[width=0.95\columnwidth]{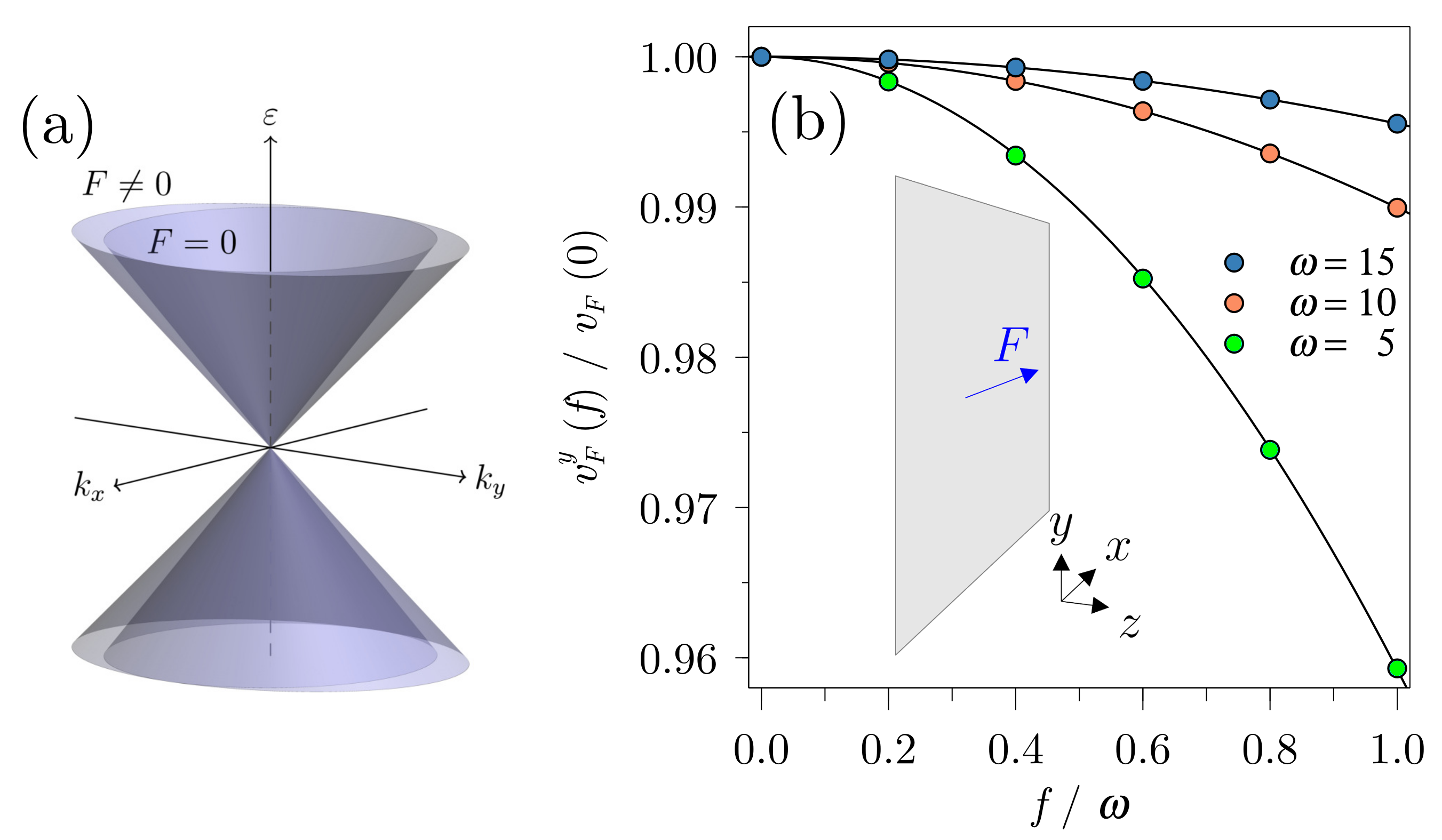}}
\caption{(a)~Dispersion relations in a topological boundary with no field and with an in-plane linearly polarized field. The Dirac cone widens anisotropically and the slope decreases quadratically with the field amplitude. Widening only occurs in the direction perpendicular to the applied field. (b)~Velocity as a function of $f/\omega$ for different values of $\omega$. Solid lines correspond to a quadratic fit of the form $1-\gamma(f/\omega)^2$, $\gamma$ being a fitting parameter.}
\label{fig:inlin}
\end{figure}

Indeed, for linearly polarized fields, the Dirac cones become anisotropic such that the cone only widens in the perpendicular direction to the field, therefore leading to an effective reduction of the velocity in that direction. This is qualitatively shown in Fig.~\ref{fig:inlin}(a) and quantitatively in Fig.~\ref{fig:inlin}(b). As a novel feature, our study establishes that in this situation the reduction of the Fermi velocity is quadratic as a function of the field magnitude, as mentioned previously. Remarkably a similar result was obtained in the context of static, crossed electric and magnetic fields~\cite{Diaz-Fernandez18}.  

\begin{figure}[htb]
\centerline{\includegraphics[width=\columnwidth]{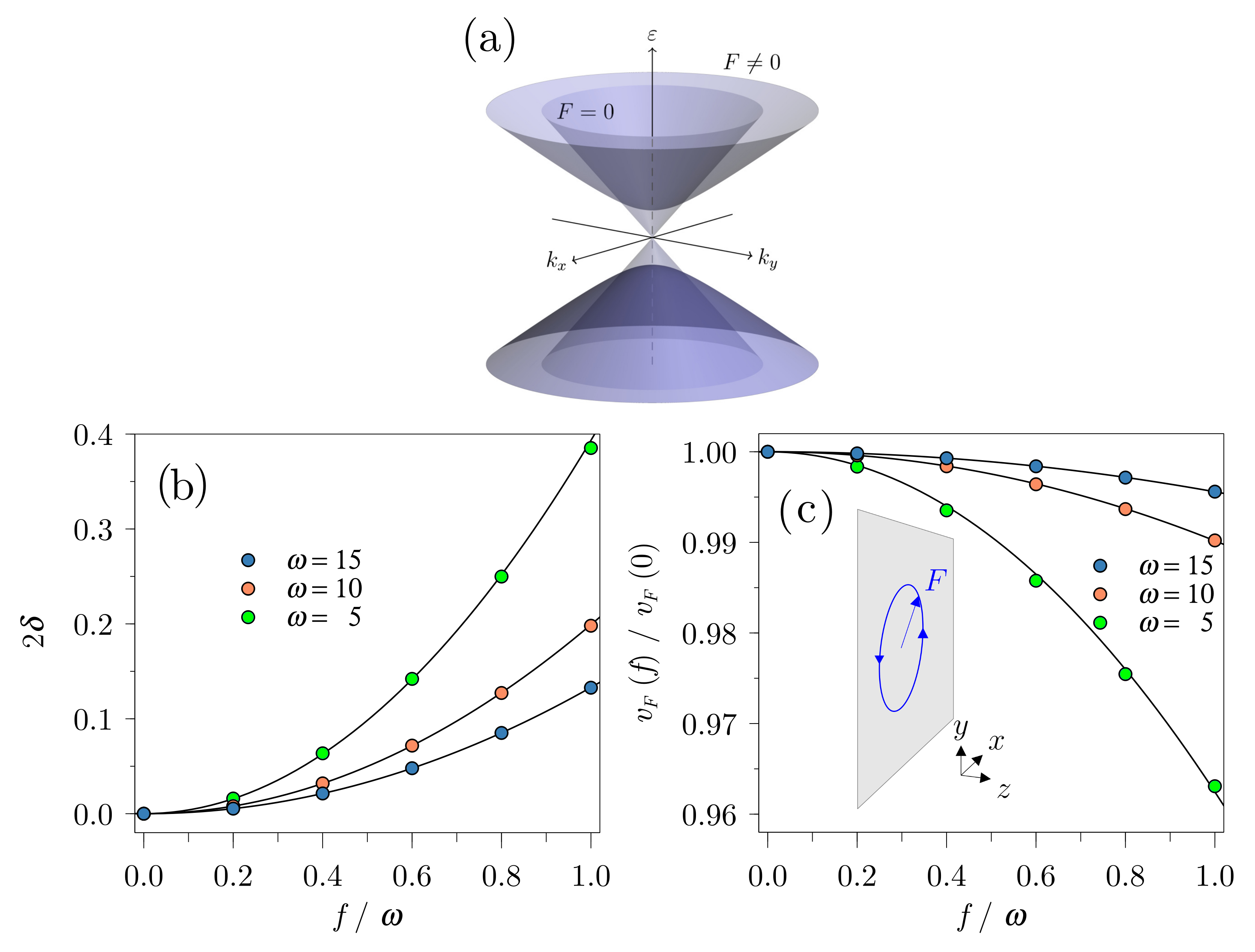}}
\caption{(a)~Dispersion relations in a topological boundary with no field and with an in-plane circularly polarized field. A gap opens up and the massive Dirac spectrum widens isotropically, the slope decreasing quadratically with the field amplitude. (c)~Gap as a function of $f/\omega$ for different values of $\omega$. Solid lines correspond to a quadratic fit of the form $\lambda(f/\omega)^2$. (c)~Velocity as a function of $f/\omega$ for different values of $\omega$. Solid lines correspond to a quadratic fit of the form $1-\gamma(f/\omega)^2$. $\lambda$ and $\gamma$ are two fitting parameters.}
\label{fig:incir}
\end{figure}

If the in-plane field is circularly polarized, a gap $2\delta$ opens up in agreement with previous studies~\cite{Kitagawa11,Wang13}, as shown qualitatively in Fig.~\ref{fig:incir}(a). Furthermore our results provides a quantitative description of how is the variation of the gap and the Fermi velocity as a function of $f/\omega$, see Fig.~\ref{fig:incir}(b) and (c). Notice that Fig.~\ref{fig:incir}(c) also shows that the resulting double-sheeted hyperboloid widens isotropically. Indeed, both the velocity and the gap can be fitted to quadratic power laws of the form $1-\gamma(f/\omega)^2$ and $\lambda(f/\omega)^2$, respectively, with $\gamma$ and $\lambda$ two fitting coefficients. 
%
%

\begin{figure}[!ht]
\centerline{\includegraphics[width=0.75\columnwidth]{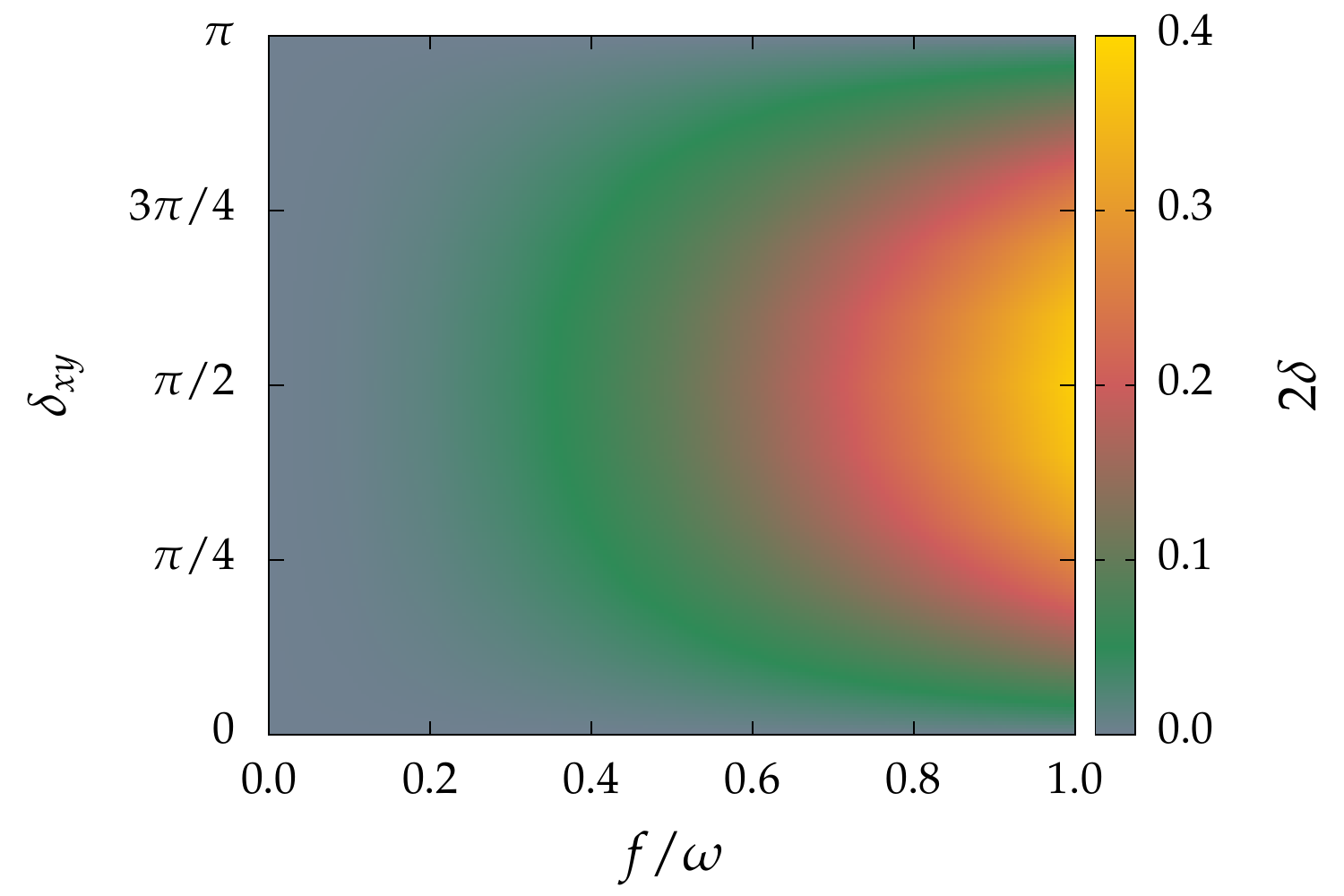}}
\caption{Density plot of the energy gap, $2\delta$, as a function of the ratio $f/\omega$ and the dephasing $\delta_{xy}$ between the $x$ and $y$ components for an in-plane configuration in a topological boundary for $\omega = 5$.}
\label{fig:density}
\end{figure}

More importantly we demonstrate that the gap that opens up can be tuned by modifying the relative phase between the $x$ and $y$ components, $\delta\varphi_{xy}=\varphi_x-\varphi_y$ and by increasing the field, as seen in Fig.~\ref{fig:density}. Indeed, if the phase difference is set to zero or $\pi$ the gap must close, but even the smallest non-zero phase difference breaks time-reversal symmetry and a gap opens up. Therefore, it is expected that the maximum gap will be at $\delta\varphi_{xy}=\pi/2$ and will increase with the field. 

\subsection{Out-of-plane fields}

\begin{figure}[htb]
\centerline{\includegraphics[width=0.95\columnwidth]{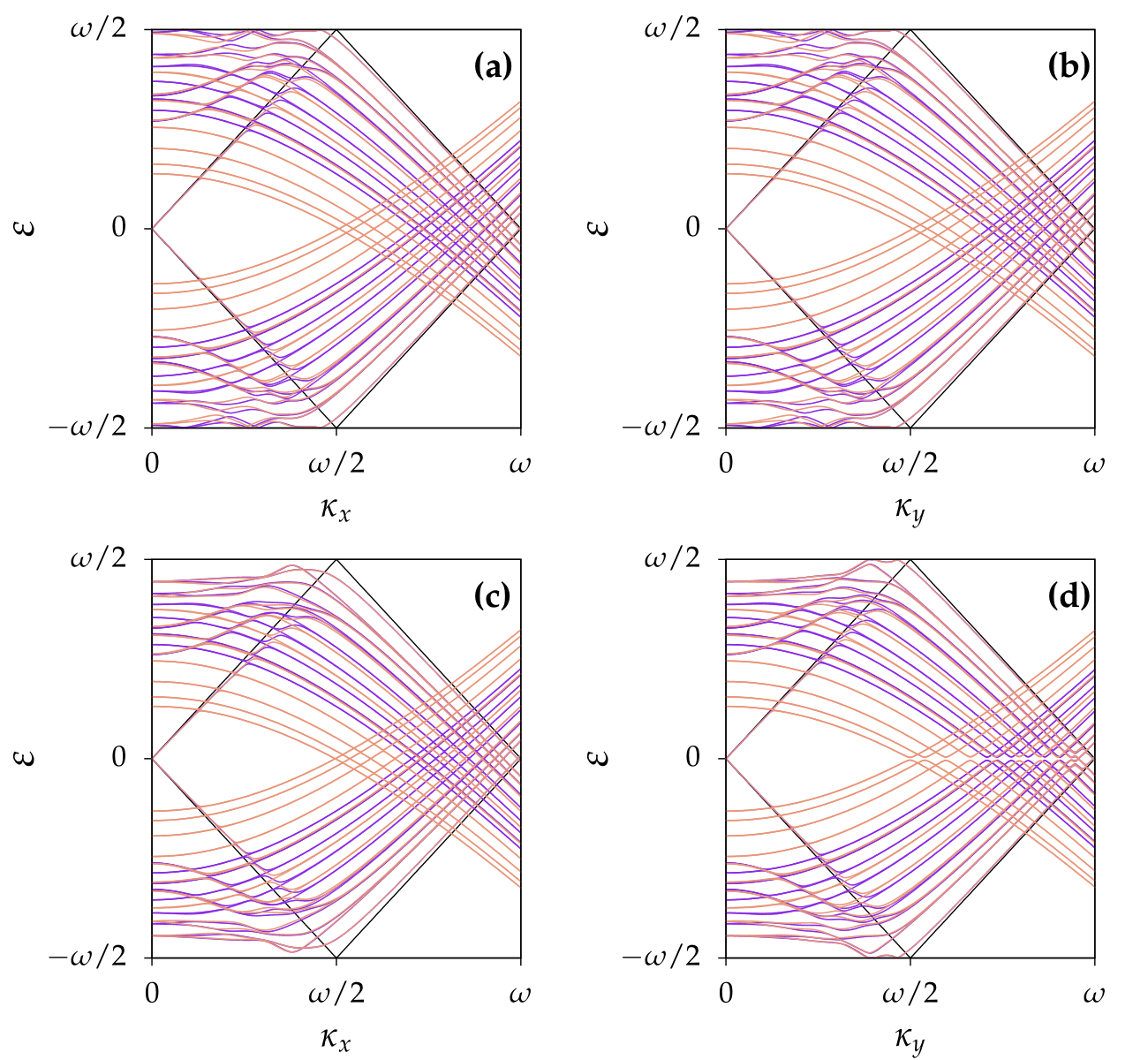}}
\caption{Quasienergy spectra for out-of-plane fields as a function of the momentum $\kappa_x$ and $\kappa_y$. In all cases, $\omega=4$ and $f=2$. Black lines indicate the Dirac cone replicas in the absence of perturbation (there would be bulk states as well). Blue and orange lines correspond to lattice spacings of $0.375$ and $0.300$, respectively. In all figures, avoided crossings occur at the Brillouin zone edges for the bulk states. Panels~(a) and~(b) correspond to linear polarization with the field along the $Z$ direction. 
Panels~(c) and~(d) correspond to circular polarization with the field contained in the $YZ$ plane. In this case, the dispersion is anisotropic, the Dirac cone widening more along the $X$ direction}
\label{fig:outspectra}
\end{figure}

Now, we now turn our discussion to those cases where there is at least one out-of-plane component of the field. In this case, we expect the Dirac point to be robust since there is no time-reversal symmetry breaking. However, as shown in Fig.~\ref{fig:outspectra}, there is hybridization with states in the bulk for large momenta. This can be understood by appealing to the static case, where hybridization is more likely to occur closer to the band edges due to proximity to the bulk states. As the number of bulk states increases due to decreasing of the lattice spacing, the avoided crossings with the Dirac state occur closer to the Dirac point. Therefore, in order to continue with the continuum description and avoid to consider the microscopic details, the spacing cannot be too small. For instance, grid spacing $0.300$ and $0.375$ correspond to small values yet sufficiently large to ignore the microscopic details. Therefore, one may argue that the Dirac cone remains unaltered for low momenta, except for a widening of the slope. This is in fact consistent with the observation that, for low momenta, the Dirac dispersion for 0.3 and 0.375 overlaps. Within these considerations, let us begin with the the detailed study of the reshaping Dirac cones under out-of-plane fields.

\begin{figure}[htb]
\centerline{\includegraphics[width=0.95\columnwidth]{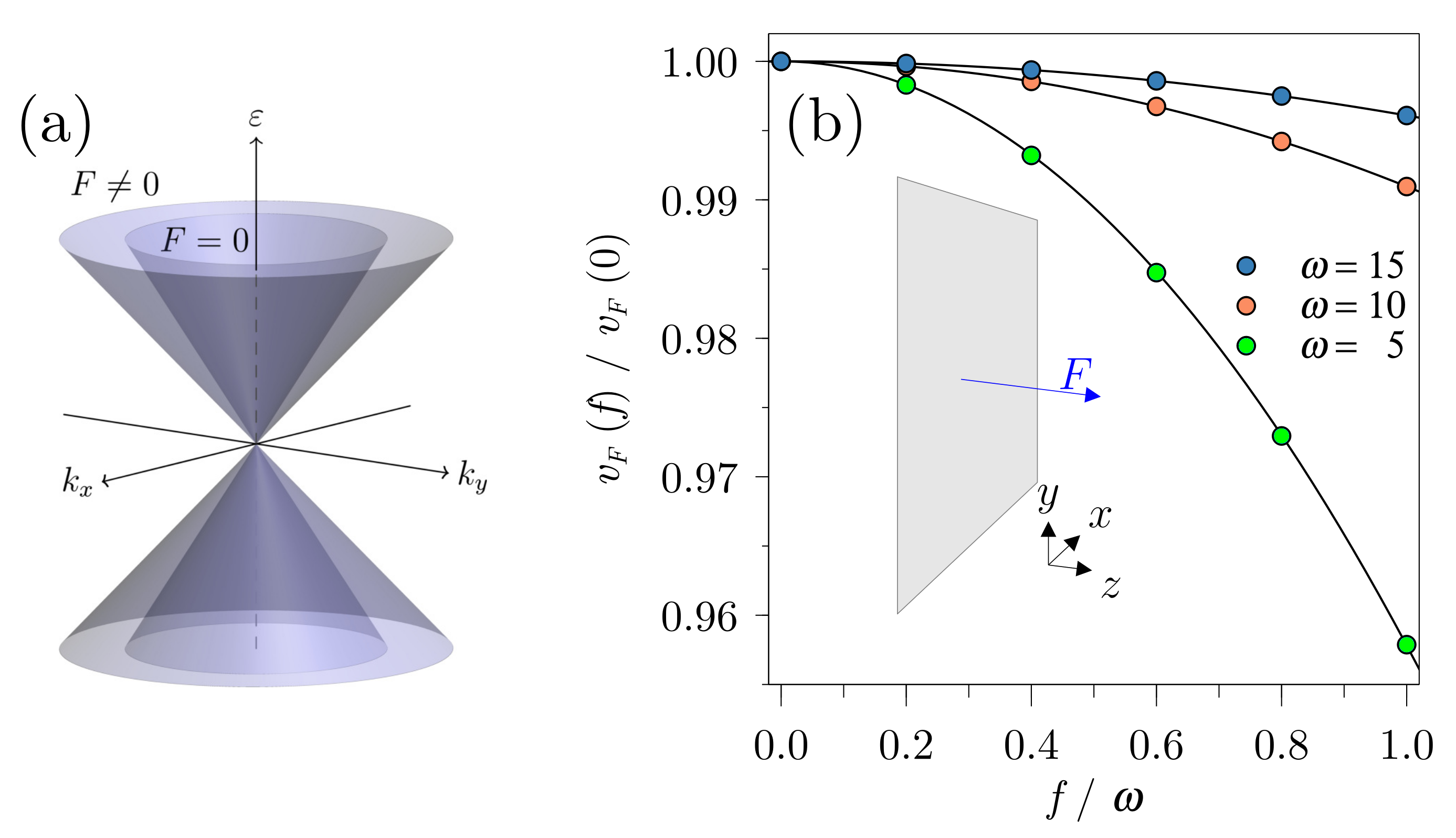}}
\caption{(a)~Dispersion relations in a topological boundary with no field and with an out-of-plane linearly polarized field. The Dirac cone widens isotropically and the slope decreases quadratically with the field amplitude. (b)~Velocity as a function of $f/\omega$ for different values of $\omega$. Solid lines correspond to a quadratic fit of the form $1-\gamma(f/\omega)^2$, $\gamma$ being a fitting parameter.}
\label{fig:outlin}
\end{figure}

\begin{figure}[htb]
\centerline{\includegraphics[width=\columnwidth]{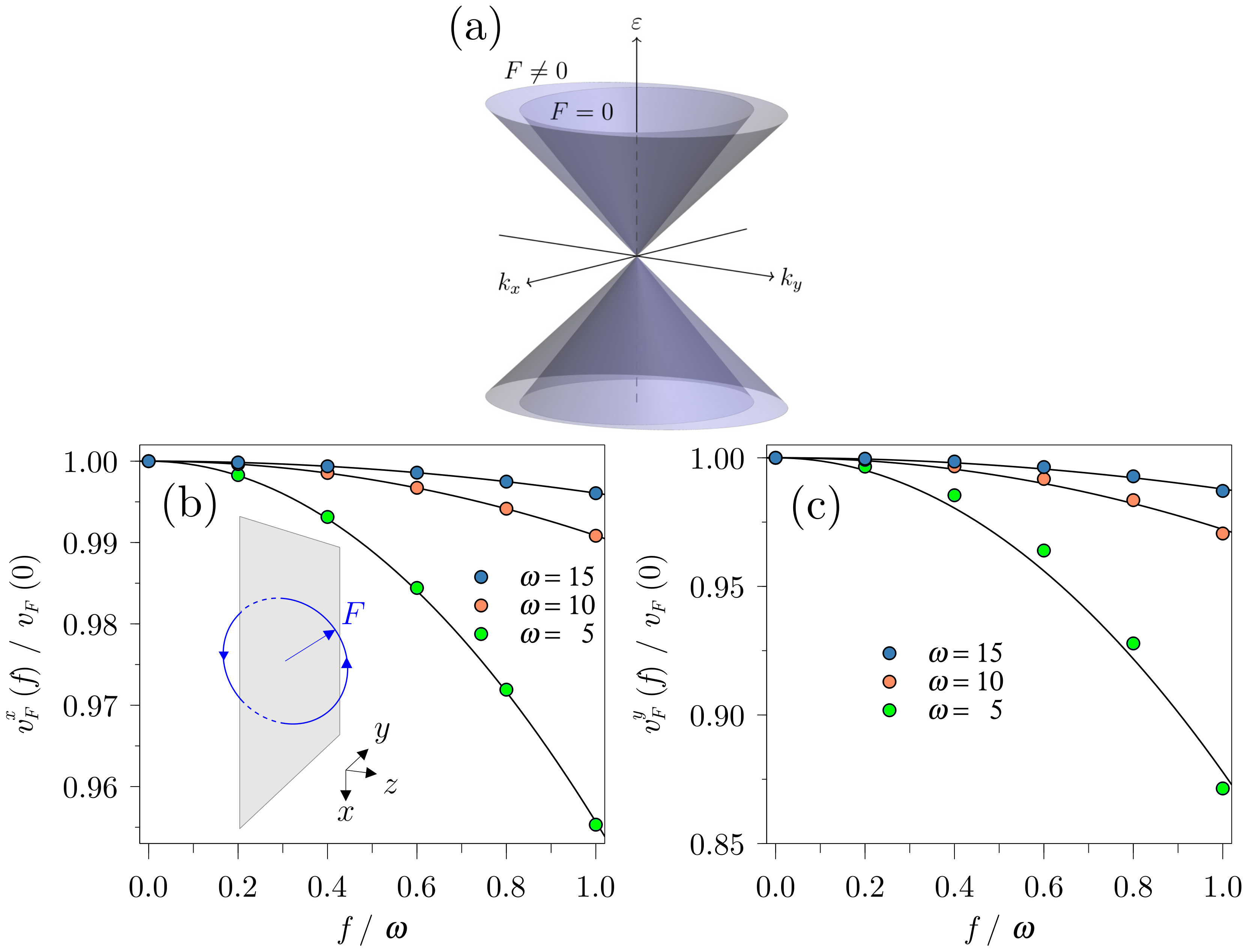}}
\caption{(a)~Dispersion relations in a topological boundary with no field and with an out-of-plane circularly polarized field. The Dirac cone widens anisotropically and the slope decreases quadratically with the field amplitude. (b)~Perpendicular and (c)~parallel velocities to the in-plane field projection as a function of $f/\omega$ for different values of $\omega$. Solid lines correspond to a quadratic fit of the form $1-\gamma(f/\omega)^2$, $\gamma$ being a fitting parameter.}
\label{fig:outcir}
\end{figure}

In the case of linearly polarized fields along the $z$-direction, the cone widens isotropically, therefore leading to an isotropic reduction of the velocity. Moreover, the velocity decreases with the field following a quadratic power law of the form $1-\gamma(f/\omega)^2$, as mentioned earlier in the text. This is displayed schematically in Fig.~\ref{fig:outlin}(a) and quantitatively in Fig.~\ref{fig:outlin}(b). Similar results were found in the static case where the field is perpendicular to the boundary~\cite{Diaz-Fernandez17a,Diaz-Fernandez17b}. 
%
%
For the circularly polarized out-of-plane field, however, we observe that the cone widens anisotropically, as shown qualitatively in Fig.~\ref{fig:outcir}(a) and quantitatively in Figs.~\ref{fig:outcir}(b) and~\ref{fig:outcir}(c). This can be explained with the results from the linearly polarized in- and out-of-plane fields. Indeed, the in-plane component leads to a reduction in the direction perpendicular to that of the field, leaving the parallel direction untouched. However, the out-of-plane component widens the cone isotropically, therefore leading to a reduction in the direction that had not been widened before and increasing the reduction in the direction that had already been affected. 
%
%

\section{Conclusions}

In this work we have shown that the Dirac cones arising at the surface of topological materials can be altered by using a periodic driving beyond previous experimental evidences. 
Remarkably within our approach we have been able to prove that some predictions based on effective surface Hamiltonians and perturbation theory\cite{Kitagawa11} are
confirmed when using a full Hamiltonian that includes bulk states.
In fact, it was known that in-plane circularly polarized light breaks time-reversal symmetry and therefore opens up a gap in the otherwise gapless Dirac cones~\cite{Wang13,Kitagawa11}. Here, we have discussed the case of a topological boundary in such a way that we can consider other configurations for the fields. Indeed, we can apply out-of-plane fields and show that the cone can widen isotropically or anisotropically, depending on the polarization. Moreover, we have observed that the reduction in the velocity squares with the applied field, a feature that was recently found also in the case of static fields~\cite{Diaz-Fernandez17a, Diaz-Fernandez17b}. Our study provides a more promising experimental set up in order to obtain an anisotropic renormalization of the velocity based on a time-periodic driving and no need for a magnetic field~\cite{Diaz-Fernandez18}. All of our findings should be straightforwardly probed by means of time- and angle-resolved photoemission spectroscopy, as discussed in Ref.~\citenum{Wang13}. We believe that our results could have an impact also in transport measurements, since a change in the velocity can lead to important reductions of the transmission. This is known to occur for other Dirac materials such as graphene on top of a patterned substrate that effectively changes the Fermi velocity~\cite{Lima16}. Using external fields, this could be achieved and lead to further control than the aforementioned setup, since the fields can be changed dynamically whereas the patterned substrate is unalterable.
 
\acknowledgments

The authors thank P. Rodr\'{i}guez for very enlightening discussions. This research has been supported by MINECO (Grants MAT2016-75955 and MAT2017-86717-P).  A.~D.-F. acknowledges support from the UCM-Santander Program (Grant CT27/16-CT28/16) and A.~G.-L. acknowledges the Juan de la Cierva program.

\section*{References}


\end{document}